\documentclass[aps,prd,superscriptaddress,nofootinbib,floats,preprint,floats,floatfix]{revtex4-1}
\usepackage{CJK}
\usepackage{bm}
\usepackage{indentfirst}
\usepackage{amsmath}
\usepackage{graphicx}
\usepackage{float}
\usepackage{amssymb}
\usepackage{subfigure}
\usepackage{amssymb}
\usepackage{psfrag}
\usepackage{hyperref}
\usepackage{tensor}
\usepackage{braket}
\usepackage{multirow}
\usepackage{verbatim}
\usepackage[dvipsnames, svgnames, x11names]{xcolor}
\hypersetup{
	colorlinks=true,
	linkcolor=red,
	citecolor=blue,
}

\allowdisplaybreaks[1]

\begin{document}
 \hfill USTC-ICTS/PCFT-23-05\\
\title{Primordial black holes and scalar induced
gravitational waves from Higgs inflation with noncanonical kinetic term}

\author{Jiong Lin}
\email{jionglin@ustc.edu.cn}
\affiliation{Interdisciplinary Center for Theoretical Study,
University of Science and Technology of China, Hefei, Anhui 230026, China}
\affiliation{Peng Huanwu Center for Fundamental Theory, Hefei, Anhui 230026, China}
\affiliation{School of Physics, Huazhong University of Science and Technology, Wuhan, Hubei
430074, China}

\author{Shengqing Gao}
\email{gaoshengqing@hust.edu.cn}
\affiliation{School of Physics, Huazhong University of Science and Technology,
Wuhan, Hubei 430074, China}

\author{Yungui Gong}
\email{yggong@hust.edu.cn}
\affiliation{School of Physics, Huazhong University of Science and Technology,
Wuhan, Hubei 430074, China}

\author{Yizhou Lu}
\email{luyz@sustech.edu.cn}
\affiliation{School of Physics, Huazhong University of Science and Technology, Wuhan, Hubei
430074, China}
\affiliation{Department of Physics, Southern University of Science and Technology, Shenzhen 518055, China}

\author{Zhongkai Wang}
\email{D202080112@hust.edu.cn}
\affiliation{School of Physics, Huazhong University of Science and Technology,
Wuhan, Hubei 430074, China}

\author{Fengge Zhang}
\email{Corresponding author.zhangfg5@mail.sysu.edu.cn}
\affiliation{School of Physics and Astronomy, Sun Yat-sen University, Zhuhai 519088, China}

\begin{abstract}
We resolve the potential-restriction problem in K/G inflation by introducing   nonminimal coupling.
In this context, Higgs field successfully drives inflation satisfying CMB observations while enhancing curvature perturbations at small scales, which in turn accounts for primordial black holes (PBHs) and scalar induced gravitational waves (SIGWs). 
We then uncover the effect of the non-canonical kinetic coupling function in more detail and study its the observational constraint.
Besides, we also give the gauge invariant expression for the integral kernel of SIGWs, which is related to terms propagating with the speed of light.
Finally, the non-Gaussian effect on PBH abundance and SIGWs is studied.  
We find that non-Gaussianity makes PBHs form more easily, 
but its effect on the energy density of SIGWs is negligible.
\end{abstract}

\maketitle

\tableofcontents

\section{Introduction}
The overdense inhomogeneities in the radiation era could gravitationally
collapse to form primordial black holes (PBHs) \cite{Carr:1974nx,Hawking:1971ei}, which could be used to account for  Dark Matters (DM) \cite{Ivanov:1994pa,Frampton:2010sw,Belotsky:2014kca,Khlopov:2004sc,
Clesse:2015wea,Carr:2016drx,Inomata:2017okj,Garcia-Bellido:2017fdg,Kovetz:2017rvv}. 
Due to the vast range of masses, PBHs may explain the
Black Hole binaries with tiny effective spin detected by LIGO-Virgo Collaboration \cite{Abbott:2016blz,Abbott:2016nmj, Bird:2016dcv,Sasaki:2016jop}. 
These density inhomogeneities can be generated from the inflationary stage, and cause collapse to form PBHs after horizon reentry.
This mechanism requires the amplitude of
the primordial curvature perturbation to be $A_s \sim \mathcal{O}(10^{-2})$ \cite{Sato-Polito:2019hws} while the amplitude has been constrained by the cosmic microwave background (CMB) anisotropy measurements to
be $A_s\approx 2.1 \times 10^{-9}$ at the pivot
scale $k_*=0.05\ \text{Mpc}^{-1}$ \cite{Akrami:2018odb}.
Thus the enhancement
of the amplitude can occur exclusively at small scales.

A way to enhance the curvature perturbation is to provide a dramatic decrease in the velocity of the inflaton, and thus the slow-roll condition is violated.
This can be achieved by an inflationary potential with an inflection point
\cite{Germani:2017bcs,Motohashi:2017kbs,Gong:2017qlj,Wu:2021zta} or a step-like feature \cite{Inomata:2021tpx}.
For other enhancement mechanisms, please see Refs. \cite{Fu:2019ttf,Kawai:2021edk,Chen:2021nio,Teimoori:2021pte,Liu:2021rgq,Zheng:2021vda,Heydari:2021gea,Heydari:2021qsr,Cai:2021wzd,Wang:2021kbh,Ahmed:2021ucx}.
While the inflection point  does lead to the decrease in $\dot{\phi}$, and thus the enhancement of the power spectrum, it is a challenge to fine-tune the model parameters to enhance the power spectrum to the order of
$\mathcal{O}(10^{-2})$ with the total number of $e$-folds
within $N\simeq 50\text{-}60$ \cite{Sasaki:2018dmp,Passaglia:2018ixg}.
Meanwhile, a new mechanism with a peak function $G(\phi)$ in the noncanonical kinetic term was proposed to enhance the primordial
power spectrum at small scales \cite{Lin:2020goi,Yi:2020kmq,Yi:2020cut,Gao:2020tsa,Yi:2021lxc,Zhang:2020uek,Zhang:2021vak,Solbi:2021wbo}.
As we will show in our paper, the peak function serves not only the enhancement of the curvature perturbation, but also the fast exit of inflation keeping the $e$-folds within $N\simeq 50\text{-}60$.
 Both sharp and broad peak functions are acceptable \cite{Yi:2020cut}, which contribute up to $\sim 20$ $e$-folds so that the usual slow-roll inflation epoch should be kept $30\text{-}40$ $e$-folds and the inflationary potential may be restricted.
To cure this problem, this mechanism was improved by generalizing the noncanonical kinetic term to $G(\phi)+f(\phi)$ \cite{Yi:2020kmq,Yi:2020cut,Gao:2020tsa,Yi:2021lxc}. 
In this paper, we will show another way to avoid this potential-restriction problem by employing the nonminimal coupling between gravity and scalar field. 

On the other hand, as the only scalar field verified so far, Higgs, if drives inflation, suffers from the problem of unacceptably large tensor-to-scalar ratio $r$. 
To satisfy CMB observation, nonminimal (derivative) couplings $\xi\phi^2R$, $G^{\mu\nu}\partial_{\mu}\phi\partial_{\nu}\phi$ are introduced to reduce $r$ \cite{Kaiser:1994vs,Bezrukov:2007ep,Germani:2010gm,Germani:2014hqa,Hamada:2014iga,Yang:2015pga,Fumagalli:2017cdo,Fumagalli:2020ody}.
Taking into account the running of self-coupling constant and the nonminimal coupling between Higgs field and gravity, the effective potential in Einstein frame possesses an inflection point, which can enhance the curvature perturbation. 
However, such an enhancement is only of five orders of magnitude compared with CMB constraint $A_s\sim \mathcal O(10^{-9})$, and is unable to produce a significant abundance of PBHs \cite{Ezquiaga:2017fvi,Bezrukov:2017dyv}.
In our paper, we will show that, by introducing a noncanonical kinetic term and a nonminimal coupling, the Higgs-field-driving inflation is compatible with CMB observation while simultaneously enhancing the curvature perturbations to order $\mathcal{O}(10^{-2})$ at small scales.

The production of PBHs by the enhanced primordial curvature perturbation
is accompanied by the generation of scalar induced gravitational waves (SIGWs)
\cite{Matarrese:1997ay,Mollerach:2003nq,Ananda:2006af,Baumann:2007zm,
Garcia-Bellido:2017aan,Saito:2008jc,Saito:2009jt,Bugaev:2009zh,
Bugaev:2010bb,Alabidi:2012ex,
Orlofsky:2016vbd,Nakama:2016gzw,Inomata:2016rbd,Cheng:2018yyr,Kawai:2021edk,Cai:2019amo,Cai:2019bmk,Cai:2019elf}, for recent review, please refer \cite{Domenech:2021ztg}, which consist of the stochastic background and can be tested by Pulsar Timing Arrays (PTA) \cite{Ferdman:2010xq,Hobbs:2009yy,McLaughlin:2013ira,Hobbs:2013aka} and the space based GW
observatories like Laser Interferometer Space Antenna (LISA) \cite{Audley:2017drz}, Taiji \cite{Hu:2017mde} and TianQin \cite{Luo:2015ght}. 
Therefore, the observations of both PBHs and SIGWs can be used to constrain the amplitude enhancement of the primordial curvature perturbation during inflation and thus to probe the physics in the early universe.

The paper is organized as follows.
In Sec. \ref{sec:2}, we show our mechanism to enhance the curvature perturbation with Higgs potential by combining the noncanonical kinetic term with the nonminimal coupling.
The PBH abundance and the energy density of SIGWs generated by Higgs inflation are presented in Sec. \ref{sec:3}.
In Sec.\ref{sec:ng}, we discuss the effect of the non-Gaussianity on PBH abundance and SIGWs.
We conclude the paper in Sec.\ref{sec:conclu}.

\section{K/G inflation with nonminimal coupling}\label{sec:2}

\subsection{Review of K/G inflation}
\begin{figure}[htp]
\centering
\includegraphics[width=0.7\textwidth]{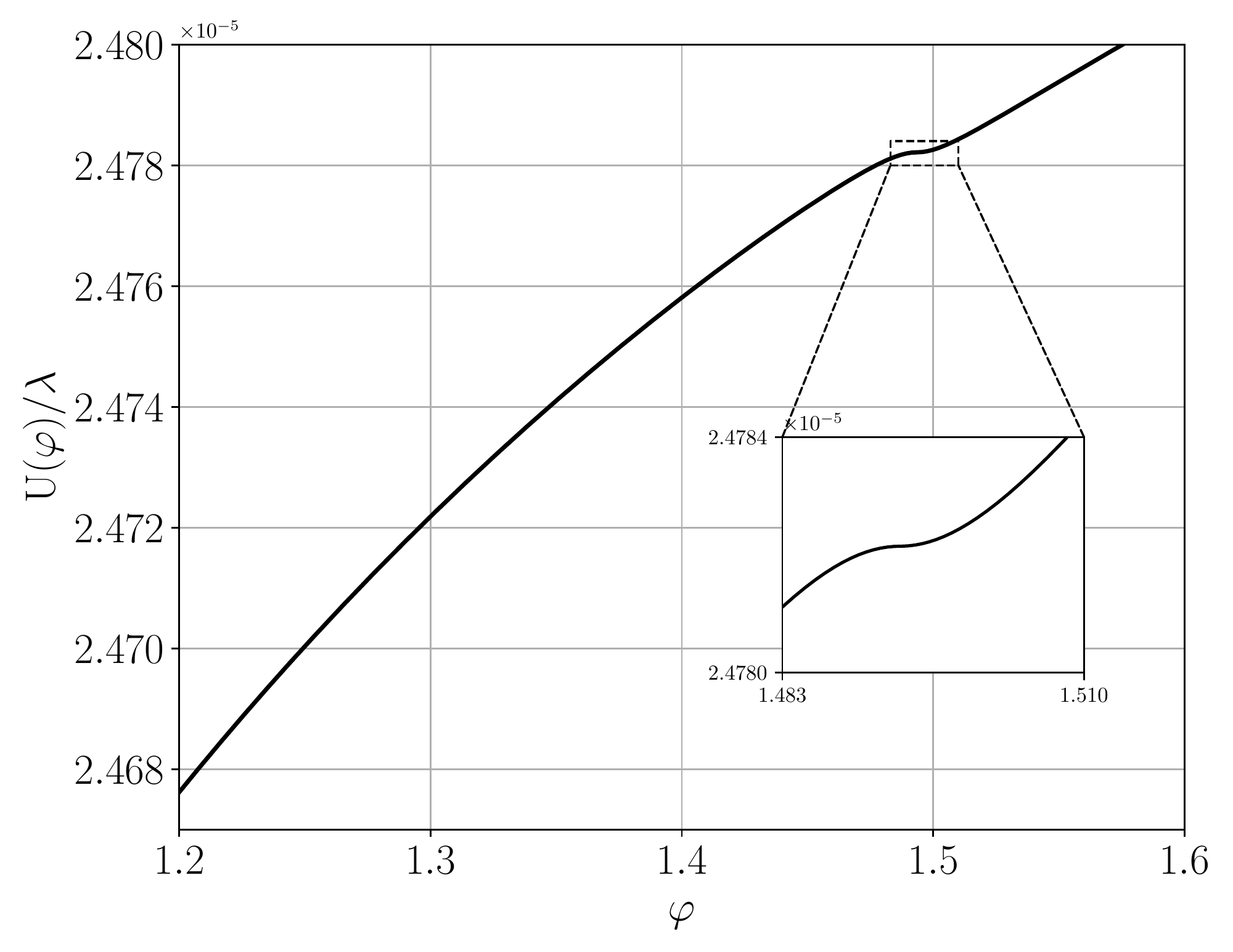}
\caption{ After the field redefinition, the corresponding canonical potential possesses an inflection point. }
\label{Fig:ep}
\end{figure}

An idea to enhance curvature perturbation is to temporarily change the friction term in curvature perturbation equation 
\begin{equation}
\label{eq:f}
\zeta''_k+2\frac{z'}{z}\zeta'_k+k^2\zeta_k=0,
\end{equation}
into a driving term during inflation.
To wit,  $z'/z<0$. 
For K/G inflation \cite{Lin:2020goi} with the noncanonical kinetic
term $[1+G(\phi)]X$ where $X=\dot{\phi}^2/2$, the friction term is
\begin{equation}
\frac{z'}{z}=aH\left[1+\epsilon_1-\epsilon_2+\frac{G_{\phi}\dot{\phi}}{2H(1+G)}\right],
\end{equation}
where $\epsilon_1=-\dot{H}/H^2, \epsilon_2=-\ddot{\phi}/(H\dot{\phi})$ are slow-roll parameters. 
To make $z'/z<0$, we could temporarily keep the second slow-roll parameter $\epsilon_2>0$ and large, i.e. the velocity of scalar field should dramatically decrease. 
The scalar field equation is
\begin{equation}
    H\dot{\phi}(3-\epsilon_2)+V_{\phi}^{\text{eff}}=0,
\end{equation}
where  
\begin{equation}
V^{\text{eff}}_{\phi}=\frac{V_{\phi}+X G_{\phi}}{1+G}
\end{equation}
is the gradient of effective potential
and the subscript $\phi$ represents the derivative with respect to $\phi$.  
In our orignal paper \cite{Lin:2020goi}, we have shown that the power spectrum can be enhanced if $G$ has a peak. Motivated by Brans-Dicke theory \cite{Brans:1961sx} with coupling $1/\phi^q$, we choose \cite{Lin:2020goi,Yi:2020cut}
\begin{equation}
\label{gfuneq1}
G(\phi)=\frac{h}{1+(\left|\phi-\phi_{c}\right|/c)^q},
\end{equation}
where $h,c$ determine the amplitude and width of the peak, respectively.
$q$ controls the shape of the enhanced power spectrum.
Larger $q$ may give a broad peak in the power spectrum.
The peak position $\phi_c$ is related to the peak mass of PBH and the peak frequency of SIGWs.
Away from the peak, $G\approx 0$ such that the usual slow-roll inflation is recovered.,
In fact, by  performing a field-redefinition $d\varphi=\sqrt{1+G(\phi)}d\phi$, K/G inflation is equivalent to a class of the canonical inflation with the  potential $U(\varphi)$ possessing an inflection point, as shown in Fig.\ref{Fig:ep}.

\subsection{Potential-restriction problem and K/G inflation with nonminimal coupling}\label{sec:2a}
Note that due to the dramatic decrease in $\dot{\phi}$, the peak function $G(\phi)$ will contribute up to $\sim 20$ $e$-folds, and the usual slow-roll inflation epoch should be kept  e-folds $N\simeq30\text{-}40$ so that the total $e$-folds during inflation is within $N\simeq 50\text{-}60$. 
Thus the usual K/G inflation suffers from the potential-restriction problem.
For power-law potential $V=\lambda\phi^p$ with $0<p\leq4$, the $e$-folds during slow-roll inflation can be expressed in terms of the spectrum index as $N_{sr}\approx(p+2)/2(1-n_s)$.
To keep $N_{sr}\simeq 30\text{-}40$, the power-law index $p$ should be bounded by $p\lesssim 1$.
Thus this mechanism does not work for Higgs field ($p=4$).  
Besides, the tensor-to-scalar ratio predicted by inflation with Higgs potential is 
\begin{equation}
    r\approx 8\left(\frac{V_{\phi}}{V}\right)^2=
    \frac{8p(1-n_s)}{p+2}\Big|_{p=4}\approx0.18,
\end{equation}
which is incompatible with observational constraints $r_{0.05}<0.036(95\% \rm{CL})$ \cite{BICEP:2021xfz}. 

To realize the enhanced power spectrum with Higgs field, we combine K/G enhancement mechanism with the nonmiminal coupling between Higgs field and gravity, i.e. $\Omega(\phi)\tilde R$ .
The action in Jordan frame is 
\begin{equation}
    S=\int d^{4} x \sqrt{-\tilde{g}}\left[\frac{1}{2} \Omega(\phi) \tilde{R}(\tilde{g})-\frac{1}{2} \omega(\phi) \tilde{g}^{\mu \nu} \nabla_{\mu} \phi \nabla_{\nu} \phi-V(\phi)\right].
\end{equation}
Under conformal transformation $g_{\mu \nu}=\Omega(\phi) \tilde{g}_{\mu \nu}$, the action in Einstein frame becomes
\begin{equation}
    S=\int d^{4} x \sqrt{-g}\left[\frac{1}{2} R(g)-\frac{1}{2} W(\phi) g^{\mu \nu} \nabla_{\mu} \phi \nabla_{\nu} \phi-U(\phi)\right],
\end{equation}
where 
\begin{equation}
    W(\phi)=\frac{3}{2} \frac{(d \Omega / d \phi)^{2}}{\Omega^{2}(\phi)}+\frac{\omega(\phi)}{\Omega(\phi)},\quad
    U(\phi)=V(\phi)/\Omega^2(\phi).
\end{equation}
For power-law potential $V(\phi)=\lambda \phi^p$, we choose the conformal factor
\begin{equation}
    \Omega(\phi)=1+\xi\phi^{p/2}.
\end{equation}
The conformal factor can flatten the power-law potential so that the tensor-to-scalar ratio is within the CMB observation and the $e$-folds of usual slow-roll inflation is kept within $30\text{-}40$. 
By choosing the appropriate coupling function $\omega(\phi)$ in Jordan frame, the coupling function in Einstein frame becomes
\begin{equation}
W(\phi)=1+G(\phi).
\end{equation}

\begin{table*}[htp]
  \centering
  \renewcommand\tabcolsep{1.8pt}
  \begin{tabular}{ccccccccccccc}
  \hline
  Models &$\phi_*$&$\phi_c$&$\lambda$&$h$&$c$&$N$&$n_s$&$r$&$k_{\text{peak}}/\text{Mpc}^{-1}$&$\mathcal{P}_{\zeta(\text{peak})}$\\
  \hline
  H1& 1.6 &1.515& $2.3\times10^{-7}$  & $8\times10^7$ & $8.3\times10^{-11}$ & 63& 0.97 &0.0007 & $2.84\times 10^{5}$ & 0.036 \\
  H2& 1.6  &1.3 & $2.3\times10^{-7}$  & $8.15\times10^7$ & $1.309\times10^{-10}$ & 63& 0.966 &0.0007 & $4\times 10^{12}$ & 0.0258 \\
  WH& 1.6  &1.47 & $2.3\times10^{-7}$  & $8\times10^7$ & $7.445\times10^{-10}$ & 64& 0.965 &0.0007 & $2.99\times 10^{5}$ & 0.016 \\
  Q& 1.3  &0.98 & $3.3\times10^{-7}$  & $1.5\times10^7$ & $9.35\times10^{-10}$ & 64& 0.965 &0.0007 & $6.53\times 10^{12}$ & 0.013 \\
  \hline
  \end{tabular}
 \caption{The chosen parameters and the results for the scalar power spectrum at small peak scales.}
\label{tab:1}
\end{table*}

\begin{figure}[htp]
\centering
\includegraphics[width=0.7\textwidth]{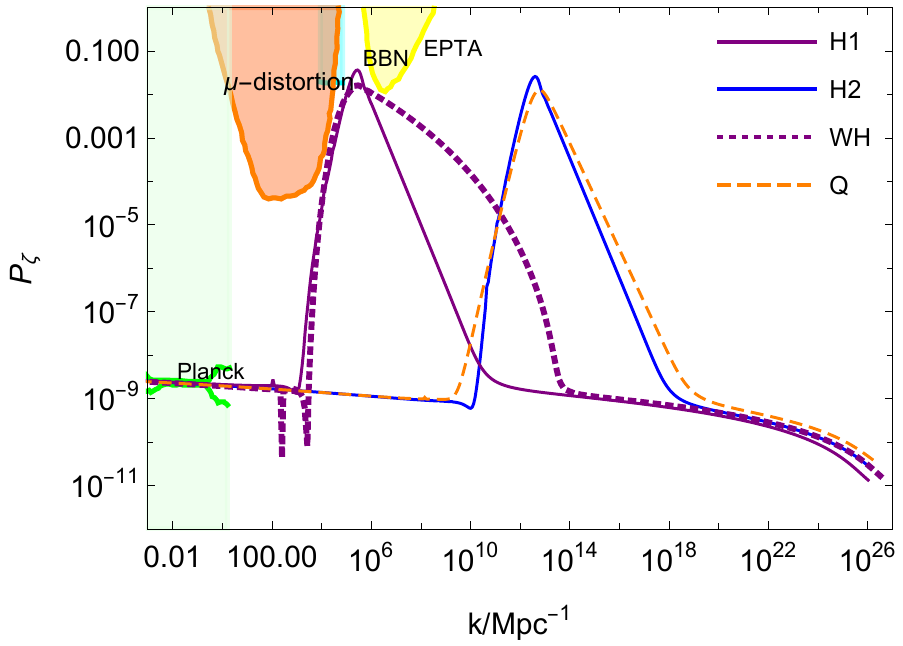}
\caption{The results for the scalar power spectrum  for
Model H1 (the purple line), Model H2 (the blue line), Model WH (the dotted purple line) and Model Q (the orange dashed line).
The green lines show the scale-dependent behavior of the power spectrum.
The light-green shaded region is excluded by the CMB observations \cite{Akrami:2018odb}. The light-grey, light-blue and thistle regions show the
constraints from $\mu$-distortion of CMB \cite{Fixsen:1996nj}, 
the effect on the ratio between neutron and proton
during the big bang nucleosynthesis (BBN) \cite{Inomata:2016uip}
and the PTA observations \cite{Inomata:2018epa}, respectively.
}
\label{Fig:1}
\end{figure}

We will show now this mechanism works for generic power-law potentials.
To be specific, we will numerically calculate the power spectrum for Higgs field ($p=4$) and power-law potential with $p=2$.
We use labels "H" and "WH" to represent Higgs inflations with the shape parameter $q=1$ and $q=6/5$, respectively and use label "Q" to represent power-law inflation with $p=2$.
The self-coupling constant $\lambda$ is set as $\mathcal{O}(10^{-7})$ to satisfy the amplitude of power spectrum  $A_s\simeq2\times10^{-9}$ at CMB scale. 
To get $\mathcal{O}(10^{-2})$ enhancement at small scale, we choose $h\sim \mathcal{O}(10^{7})$.
The nonminimal coupling constant is taken as $\xi=100$. 
With the model parameters listed in Table \ref{tab:1}, solving the equations for the background and the perturbations numerically,  the results for the scalar power spectrum are shown in Table \ref{tab:1} and Fig.\ref{Fig:1}.
From these results, we can see that $n_s$ and $r$ are well within the CMB observation constraints, $n_s=0.9649 \pm 0.0042 (68\% \rm{CL})$ and $r_{0.05} < 0.036
(95\% \rm{CL})$ \cite{Akrami:2018odb,BICEP:2021xfz}.
In particular, due to conformal factor, $r$ can be reduced to order $\mathcal{O}(10^{-4})$.
The total $e$-folds are around $60$.
The power spectrum is enhanced to $\mathcal{O}(10^{-2})$ at the scales $\mathcal{O}(10^{5})~\rm{Mpc}^{-1}$ and  $\mathcal{O}(10^{12})~\rm{Mpc}^{-1}$.
To be specific, the power spectrum for models H1 and WH are enhanced at the scale $\mathcal{O}(10^{5})~\rm{Mpc}^{-1}$ and the power spectrum for models H2 and Q are enhanced at the scale $\mathcal{O}(10^{12})~\rm{Mpc}^{-1}$.
In addition, the shape parameter $q=1$ produces a sharp peak while the larger shape parameter $q=6/5$ produces the broad peak. 

\subsection{Non-canonical kinetic coupling function and its observational constraint}

In this subsection, we will uncover the effect of non-canonical kinetic coupling function in more detail and study the observational constraint on the parameter space of non-canonical kinetic coupling function.

\begin{figure}[htp]
\centering
\includegraphics[width=0.7\textwidth]{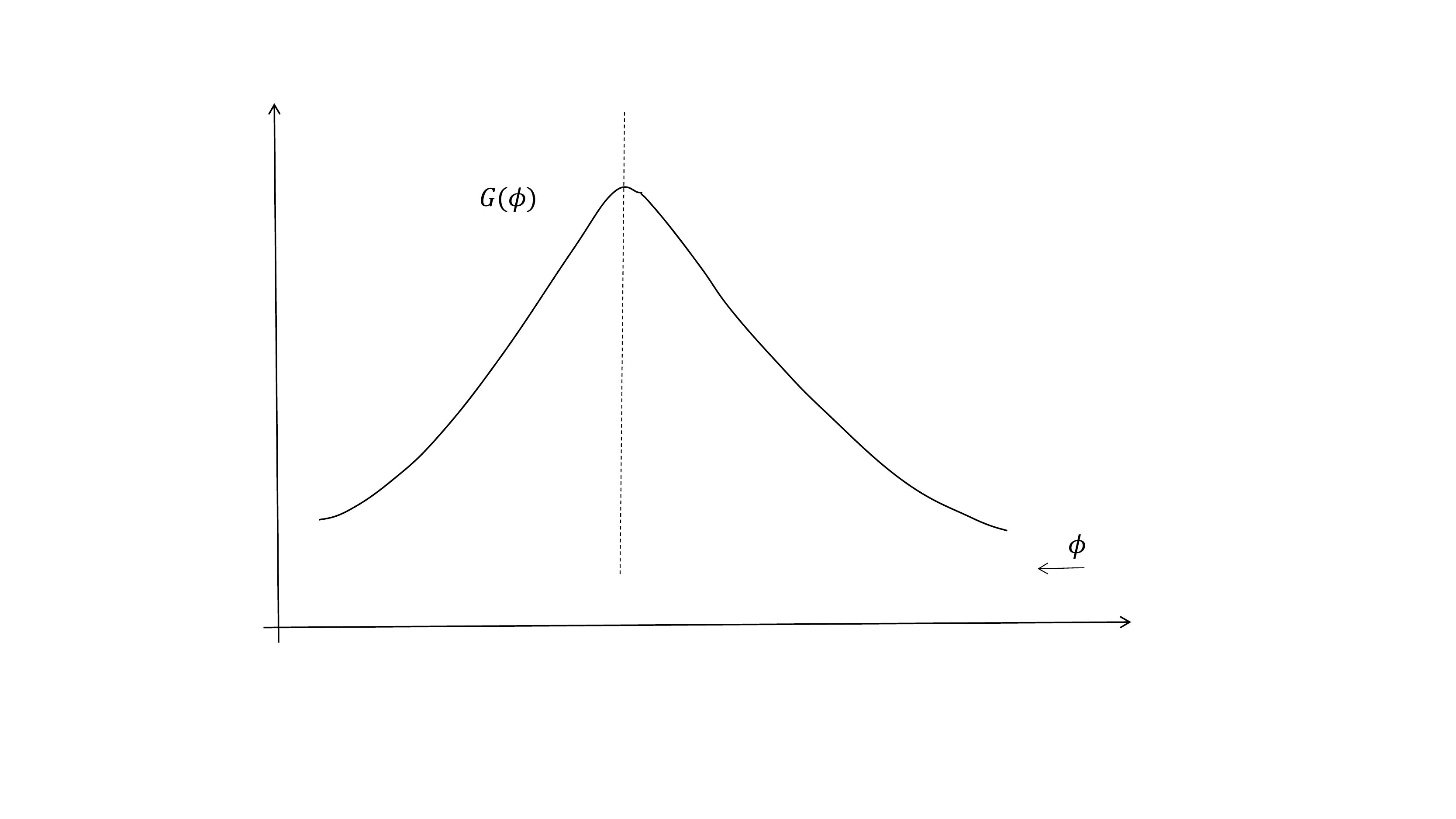}
\includegraphics[width=0.7\textwidth]{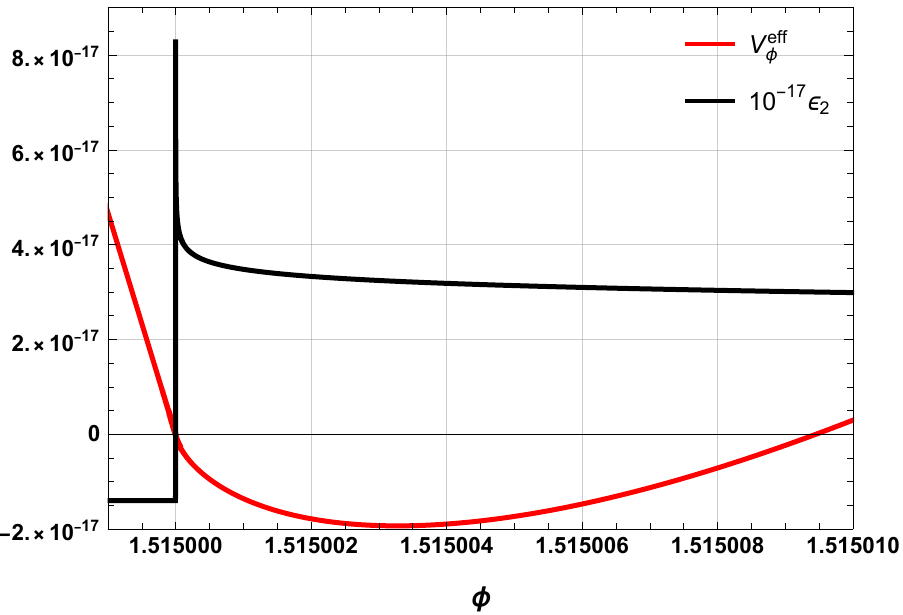}
\caption{The peak function $G(\phi)$ and 
numerical results for the behaviors of  $V^{\text{eff}}_{\phi}$ and $\epsilon_2$ around the peak $\phi_c=1.515$ for Model H1.}
\label{Fig:g}
\end{figure}

In Fig.\ref{Fig:g}, we show the numerical results for the behaviors of the gradient of effective potential $V^{\text{eff}}_{\phi}$ and the second slow-roll parameter $\epsilon_2$ around the peak $\phi_c=1.515$ for Model H1.
As we can see, as 
$\phi$ rolls down to the right region of the peak where $G_{\phi}$ is negative and very large,
the gradient of effective potential satisfies $V^{\text{eff}}_{\phi}<0$ so that $\epsilon_2>3$ and $\dot{\phi}$ will dramatically decrease, and thus enhance curvature perturbation. 
As $\phi$ leaves for the left region where  $G_{\phi}$ is positive and very large, the gradient of effective potential $V^{\text{eff}}_{\phi}>0$ such that $\epsilon_2<0$.
Here $\dot{\phi}$ will violently increase and help  inflation exit.
To sum up, phenomenologically, the peak function $G$ enables not only the dramatic decrease in $\dot{\phi}$, which further leads to the enhancement of the curvature perturbation but also the fast exit of inflation.

\begin{table*}[htp]
  \centering
  \begin{tabular}{cccccccc}
  \hline
 $c$ &$8.3\times10^{-11}$&$8.8\times10^{-11}$&$9.3\times10^{-11}$&$1\times10^{-10}$&$2\times10^{-10}$\\
  \hline
 Upper bound on $h$  &$8.05\times10^{7}$ & $7.6\times10^7$ & $7.2\times 10^7$& $6.7\times10^7$ & $3.4\times10^7$\\
  \hline
  \end{tabular}
 \caption{Constraint on K/G model with with Higgs potantial from LIGO merger rate.
 Here $\phi_c=1.515$ and $q=1$.}
\label{tab:11}
\end{table*}

\begin{table*}[htp]
  \centering
  \begin{tabular}{cccccccc}
  \hline
 $c$ &$1.4\times10^{-10}$&$2\times10^{-10}$&$2.5\times10^{-10}$&$3\times10^{-10}$&$3.2\times10^{-10}$\\
  \hline
Upper bound on $h$ &$8\times10^{7}$ & $5.65\times10^7$ & $4.55\times 10^7$& $3.85\times10^7$ & $3.7\times10^7$\\
  \hline
  \end{tabular}
 \caption{Constraint on K/G model with with Higgs potential from White Dwarf Explosion. 
 Here $\phi_c=1.28$ and $q=1$.}
\label{tab:12}
\end{table*}

The observational constraints on the present PBH abundance can be used to to constrain the power spectrum for primordial curvature perturbations at small scales and thus the range of parameter space of non-canonical kinetic coupling function.
Now let us study the observation constraints on K/G model with Higgs potential in detail.
The LIGO merger rates \cite{Ali-Haimoud:2017rtz} constrain the power spectrum as $\mathcal{P}_{\zeta}\lesssim 0.04$ for 
$8.9\times10^4~\text{Mpc}^{-1}\lesssim k\lesssim 4.9\times10^5~\text{Mpc}^{-1}$ \cite{Sato-Polito:2019hws}.
This $k$-range of this constraint can be used to bound the amplitude parameter $h$ of our models with the peak position $1.5\lesssim \phi_c$\footnote{The amplitude parameter $h$ with a larger $\phi_c$ is bounded by LIGO merger rates constrains, as for a larger $\phi_c$, one could always choose a larger $h$ so that the corresponding peak scale locates within the scale where LIGO merger rates constrains.} for the shape factor $q=1$ and $1.47\lesssim \phi_c$ for $q=6/5$, which could further apply to the realizations of Models H1 and WH.
To be specific, by choosing the peak position $\phi_c=1.515$ and $q=1$ and varying the width parameter $c$, we numerically find the upper bound on the amplitude parameter $h$  and the results are shown in Tab.\ref{tab:11}.
The larger the parameter $c$ is, the lower the upper bound on $h$ will be.
This result is well comprehensive.
For a wider peak function, the velocity of the inflaton decreases more dramatically. Therefore a smaller amplitude parameter $h$ is required to realize the same enhancement on the curvature perturbation.
Moreover, the White Dwarf Explosion \cite{Graham:2015apa} constrains the power spectrum to be $\mathcal{P}_{\zeta}\lesssim 0.023$ for $5.1\times10^{12}\text{Mpc}^{-1}\lesssim k\lesssim 2\times10^{13}\text{Mpc}^{-1}$.
This $k$-range of this constraint can be used to bound the amplitude parameter $h$ of our models with the peak position $1.27\lesssim \phi_c $ for $q=1$, which could further apply to the realizations of Models H2.
The corresponding bound on $h$ is shown in Tab.\ref{tab:12} and here we choose $\phi_c=1.28$ and $q=1$.

\section{Primordial black holes and scalar induced
gravitational waves}\label{sec:3}
The large curvature perturbation from inflation can induce PBHs and GWs at radiation era. 
In this section, we will calculate PBHs abundance and SIGWs from K/G inflation with nonminimal coupling.
Before that, we will first consider the gauge issue on SIGWs and give a gauge invariant expression for the integral kernel of SIGWs.

\subsection{The gauge invariant expression for the integral kernel of SIGWs}
Considering a metric perturbation
\begin{equation}
\mathrm{d} s^{2}=a^{2}\left[-(1+2 \phi) \mathrm{d} \tau^{2}+2 B_{, i} \mathrm{~d} x^{i} \mathrm{~d} \tau+\left((1-2 \psi) \delta_{i j}+2 E_{, i j}+\frac{1}{2} h_{i j}^{\mathrm{TT}}\right) \mathrm{d} x^{i} \mathrm{~d} x^{j}\right],
\end{equation}
the scalar-induced tensor perturbations $h^{\text{TT}}_{ij}$ satisfy
\cite{Ananda:2006af,Baumann:2007zm}
\begin{align}
\label{theeq}
  h_{ij}^{\mathrm{TT}\prime\prime}+2\mathcal{H}h_{ij}^{\mathrm{TT}\prime}-\nabla^2h_{ij}^{\mathrm{TT}}=4\mathcal{T}_{ij}^{lm}s_{lm},
\end{align}
where $\mathcal{H}= aH$,
$\mathcal{T}_{ij}^{lm}$ is the projection tensor extracting the transverse and traceless part of a tensor
and the scalar source is \cite{Lu:2020diy}
\begin{equation}
\label{source}
\begin{split}
 -s_{ij}=&\psi_{,i}\psi_{,j}+\phi_{,i}\phi_{,j}
 -\sigma_{,ij}\left(\phi^\prime+\psi^\prime-\nabla^2\sigma\right)
 +\left(\psi_{,i}^\prime\sigma_{,j}+\psi_{,j}^\prime\sigma_{,i}\right)
 -\sigma_{,ik}\sigma_{,jk}+2\psi_{,ij}\left(\phi+\psi\right)\\
 &-8\pi Ga^2({\rho_0}+{P_0})\delta V_{,i}\delta V_{,j}-2\psi_{,ij}\nabla^2E
 +2E_{,ij}\left(\psi^{\prime\prime}+2\mathcal{H}\psi^\prime-\nabla^2\psi\right)-E_{,ik}^\prime E_{,jk}^\prime\\
 &+E_{,ikl}E_{,jkl}+2\left(\psi_{,jk}E_{,ik}+\psi_{,ik}E_{,jk}\right)
 -2\mathcal{H}(\psi_{,i}E_{,j}^\prime+\psi_{,j}E_{,i}^\prime)-\left(\psi_{,i}^\prime E_{,j}^\prime+\psi_{,j}^\prime E_{,i}^\prime\right)\\
 &
 -\left(\psi_{,i}E_{,j}^{\prime\prime}+\psi_{,j}E_{,i}^{\prime\prime}\right)
 +2E_{,ij}^\prime\psi^\prime+E_{,ijk}\left(E^{\prime\prime}+2\mathcal{H}E^\prime-
 \nabla^2E\right)_{,k},
 \end{split}
 \end{equation}
where $\sigma= E^\prime-B$ is the shear potential.
Solving eq. \eqref{theeq}, the current energy density of SIGWs can be expressed as  \cite{Inomata:2016rbd,Kohri:2018awv}
\begin{equation}
\label{gwres1}
\begin{split}
\Omega_{\mathrm{GW}}(k,\eta_0)=&\frac{1}{6}\frac{\Omega_r(\eta_0)}{\Omega_r(\eta)}\left(\frac{k}{aH}\right)^2
\int_{0}^{\infty}dv\int_{|1-v|}^{1+v}du\left\{
\left[\frac{4v^2-(1-u^2+v^2)^2}{4uv}\right]^2\right.\\
&\left. \times \overline{I_{\text{RD}}^{2}(u, v, x\to \infty)} \mathcal{P}_{\zeta}(kv)\mathcal{P}_{\zeta}(ku)\right\},
\end{split}
\end{equation}
where $u=|\bm{k}-\tilde{\bm{k}}|/k$, $v=\tilde{k}/k$, $x=k\tau$,
 $\Omega_r$ is the fraction energy density of radiation, the overbar denotes the oscillation time average and $I_{\text{RD}}$ is the integral kernel in radiation domination.
 In Newtonian gauge, the integral kernel is
 \begin{equation}
\label{irdeq1}
\begin{split}
I_{\text{N}}(u,v,x)=&\int_1^x dy\, y \sin(x-y)\{3T_N(uy)T_N(vy)\\
&+y[T_N(vy)uT_N'(uy)+vT_N'(vy)T_N(uy)]
+y^2 u v T_N'(uy)T_N'(vy)\},
\end{split}
\end{equation}
where the transfer function $T_N$ in the radiation domination is
\begin{equation}
\label{transfer}
T_N(x)=\frac{9}{x^2}\left(\frac{\sin(x/\sqrt{3})}{x/\sqrt{3}}-\cos(x/\sqrt{3})\right).
\end{equation}
The analytical expression for $I_{\text{RD}}$ in Newtonian gauge was given in Refs. \cite{Espinosa:2018eve,Lu:2019sti,Kohri:2018awv}.
However, SIGWs suffer from the gauge issue \cite{Hwang:2017oxa,Tomikawa:2019tvi,Lu:2020diy,Ali:2020sfw,DeLuca:2019ufz,Inomata:2019yww,Yuan:2019fwv,Domenech:2020xin,Chang:2020iji,Cai:2021ndu}.
On one hand, this issue may be related to the definitions of gravitational waves and their energy \cite{Cai:2021ndu}.
On the other hand, as noted in Ref. \cite{Inomata:2019yww,Ali:2020sfw}, only terms that oscillate as $\sin x$ and $\cos x$ propagate with the speed of light.
They are taken as genuine GWs out of second-order tensor perturbations.
Now let us write down the integral kernel in arbitrary gauge and extract the terms that propagate with speed of light.
The integral kernel in arbitrary gauge can be obtained by the transformation
\begin{equation}
I_{\mathrm{N}}(u, v, x) \rightarrow I_{\mathrm{N}}(u, v, x)+I_{\chi}(u, v, x),
\end{equation}
where
\begin{align}
\label{SIGW_ichi_newton}
   I_\chi(u,v, x)= &- \frac{1}{4uv}\left(\frac{3+3w}{5+3w}\right)^2\left[\vphantom{\frac{u^2}{v^2}}
   -4\left(\frac{u}{v}T_\mathrm{N}(u x)T_\beta(v x)
   +\frac{v}{u}T_\mathrm{N}(v x)T_\beta(u x)\right)\right.\notag
   \\
   &+2T_\alpha(u x)T_\alpha(v x)+4\frac{\mathcal{H}}{k}
   \left(\frac1v T_\alpha(u x)T_\beta(v x)+\frac1u T_\beta(u x)T_\alpha(v x)\right)\notag
   \\&
	\left.+\frac{1-u^2-v^2}{uv}T_\beta(u x)T_\beta(v x)\right],
\end{align}
$T_{\alpha},T_{\beta}$ are related to gauge transformation. Note that the gauge transformation can be expressed in terms of scalar perturbations, thus $I_{\chi}$  only contains terms with sound speed $c_s^2=w$ instead of the speed of light. 
Omitting terms that do not propagate with speed of light, the kernel of genuine SIGWs (the terms of $\sin x$ and $\cos x$, which propagate with the speed of light) in arbitrary gauge is\footnote{This is the solution with the lower limit being 0 in \eqref{irdeq1}.
In fact, according to Ref. \cite{Lu:2019sti}, the difference can usually be ignored.
}
\begin{align}\label{SIGW_SIGW_GI}
\quad &I_\mathrm{GW}(u,v, x)\notag\\
=&\frac{3}{4u^3v^3 x}\Bigg\{
-4uv(u^2+v^2-3)\sin x+(u^2+v^2-3)^2\notag\\
&\times\bigg[
\sin x\bigg(
\mathrm{Ci}\left[\left(1+\frac{u-v}{\sqrt{3}}\right) x\right]+\mathrm{Ci}\left[\left(1+\frac{v-u}{\sqrt{3}}\right) x\right]-\mathrm{Ci}\left[\left(1+\frac{u+v}{\sqrt{3}}\right) x\right]\notag\\&
-\mathrm{Ci}\left[\left|
1-\frac{u+v}{\sqrt{3}}
\right|x\right]
+\log\left|
\frac{3-(u+v)^2}{3-(u-v)^2}
\right|
\bigg)+\cos x\bigg(-\mathrm{Si}\left[\left(1+\frac{u-v}{\sqrt{3}}\right) x\right]\notag\\
&
-\mathrm{Si}\left[
  \left(1+\frac{v-u}{\sqrt{3}}\right)x
\right]+\mathrm{Si}\left[\left(1-\frac{u+v}{\sqrt{3}}\right)x\right]
+\mathrm{Si}\left[\left(1+\frac{u+v}{\sqrt{3}}\right)x\right]
\bigg)
\bigg]
\Bigg\},
\end{align}
where
\begin{equation}
\begin{aligned}
&\operatorname{Si}(x) \equiv \int_{0}^{x} \frac{\sin y}{y} \mathrm{~d} y, \ \operatorname{Ci}(x) \equiv-\int_{x}^{\infty} \frac{\cos y}{y} \mathrm{~d} y .
\end{aligned}
\end{equation}
At late time $x\to\infty$, the integral kernel of genuine SIGWs becomes
\begin{equation}
\begin{aligned}
\overline{I_{\mathrm{GW}}^{2}(u, v, x\rightarrow \infty)}
=&\overline{I_{\mathrm{N}}^{2}(u, v, x\rightarrow \infty)}\\
=& \frac{1}{2 x^{2}}\left(\frac{3\left(u^{2}+v^{2}-3\right)}{4 u^{3} v^{3}}\right)^{2}\left\{\left(-4 u v+\left(u^{2}+v^{2}-3\right) \log \left|\frac{3-(u+v)^{2}}{3-(u-v)^{2}}\right|\right)^{2}\right.\\
&\left.+\pi^{2}\left(u^{2}+v^{2}-3\right)^{2} \Theta(u+v-\sqrt{3})\right\},
\end{aligned}
\end{equation}
where the HeavisideTheta function 
\begin{equation}
\Theta(x)= \begin{cases}1, & x \geqslant 0 \\ 0, & x<0\end{cases}.
\end{equation}

\subsection{PBHs and SIGWs from Higgs inflation}
The overdense region would gravitationally collapse to form PBHs when horizon reentry during radiation dominated era.
The current fractional energy density of PBHs with mass $M$ to DM is \cite{Carr:2016drx,Gong:2017qlj}
\begin{equation}
\label{fpbheq1}
\begin{split}
Y_{\text{PBH}}(M)=&\frac{\beta(M)}{3.94\times10^{-9}}\left(\frac{\gamma}{0.2}\right)^{1/2}
\left(\frac{g_*}{10.75}\right)^{-1/4}\left(\frac{0.12}{\Omega_{\text{DM}}h^2}\right)
\left(\frac{M}{M_\odot}\right)^{-1/2},
\end{split}
\end{equation}
where $M_{\odot}$ is the solar mass, $\gamma= 0.2$  \cite{Carr:1975qj}.
$g_*$ is the effective degrees of freedom at the formation time.
For the temperature $T>300$ GeV, $g_*=107.5$ and for $0.5\ \text{MeV}<T<300\ \text{GeV}$, $g_*=10.75$.
$\Omega_{\text{DM}}$ is the current
energy density parameter of DM and we take $\Omega_{\text{DM}}h^2=0.12$ \cite{Aghanim:2018eyx}.
 The PBH mass $M$ is related to the scale $k$ as \cite{Gong:2017qlj}
\begin{equation}
\label{mkeq1}
M(k)=3.68\left(\frac{\gamma}{0.2}\right)\left(\frac{g_*}{10.75}\right)^{-1/6}
\left(\frac{k}{10^6\ \text{Mpc}^{-1}}\right)^{-2} M_{\odot}.
\end{equation}
$\beta(M)$ is the fractional energy density of PBHs
at the formation. 
For Gaussian comoving curvature perturbation $\zeta$ \cite{Ozsoy:2018flq,Tada:2019amh}, we have
\begin{equation}\label{betaG}
    \beta^G(M) \approx \sqrt{\frac{2}{\pi}}\frac{\sqrt{\mathcal{P}_{\zeta}}}{\mu_c}
\exp\left(-\frac{\mu_c^2}{2\mathcal{P}_{\zeta}}\right),
\end{equation} 
where $\delta_c$ is the threshold for the PBH formation and $\mu_c=9\sqrt{2}\delta_c/4 $ \cite{Yi:2020cut}.
Here we choose $\delta_c=0.4$ \cite{Musco:2012au,Harada:2013epa,Tada:2019amh,Escriva:2019phb,Yoo:2020lmg} for calculations.

\begin{figure}[htp]
\centering
\includegraphics[width=0.7\textwidth]{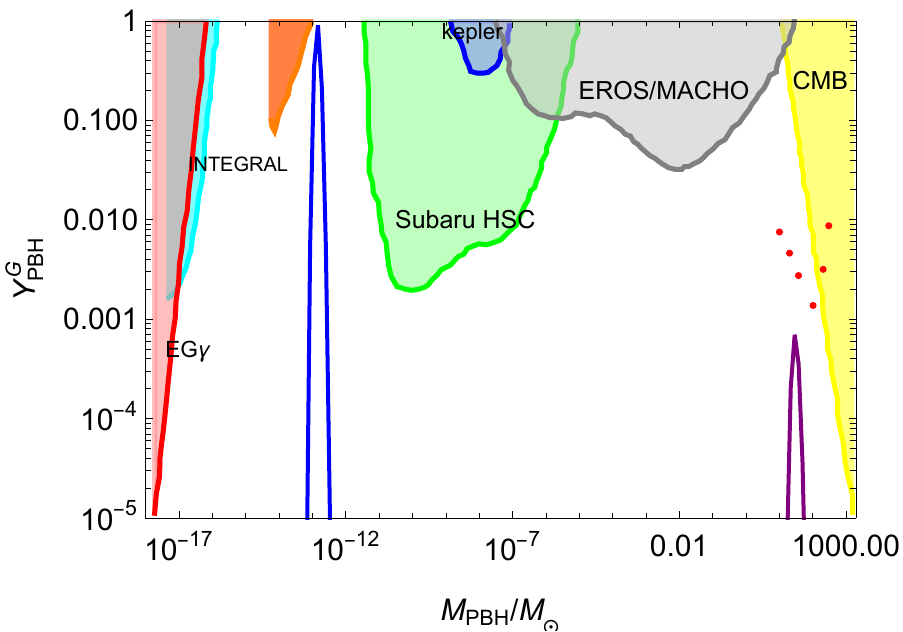}
\caption{The PBH abundances for
Model H1 (the purple line) and Model H2 (the blue line) using the Gaussian formulate $\beta^G$ for the fraction energy density of  PBHs at the formation.
The shaded regions show the observational constraints on the PBH abundance:the red region from extragalactic gamma-rays by PBH evaporation (EG$\gamma$) \cite{Carr:2009jm},
the cyan region from galactic center 511 keV gamma-ray line (INTEGRAL) \cite{Laha:2019ssq,Dasgupta:2019cae},
the orange region from white dwarf explosion (WD) \cite{Graham:2015apa},
the green region from microlensing events with Subaru HSC \cite{Niikura:2017zjd},
the blue region from the Kepler satellite \cite{Griest:2013esa},
the gray region from the EROS/MACHO \cite{Tisserand:2006zx},
the red points from LIGO merger rate \cite{Ali-Haimoud:2017rtz}
and the yellow region from accretion constraints by CMB \cite{Ali-Haimoud:2016mbv,Poulin:2017bwe}.
}
\label{Fig:pbh}
\end{figure}

\begin{figure}[htp]
\centering
\includegraphics[width=0.7\textwidth]{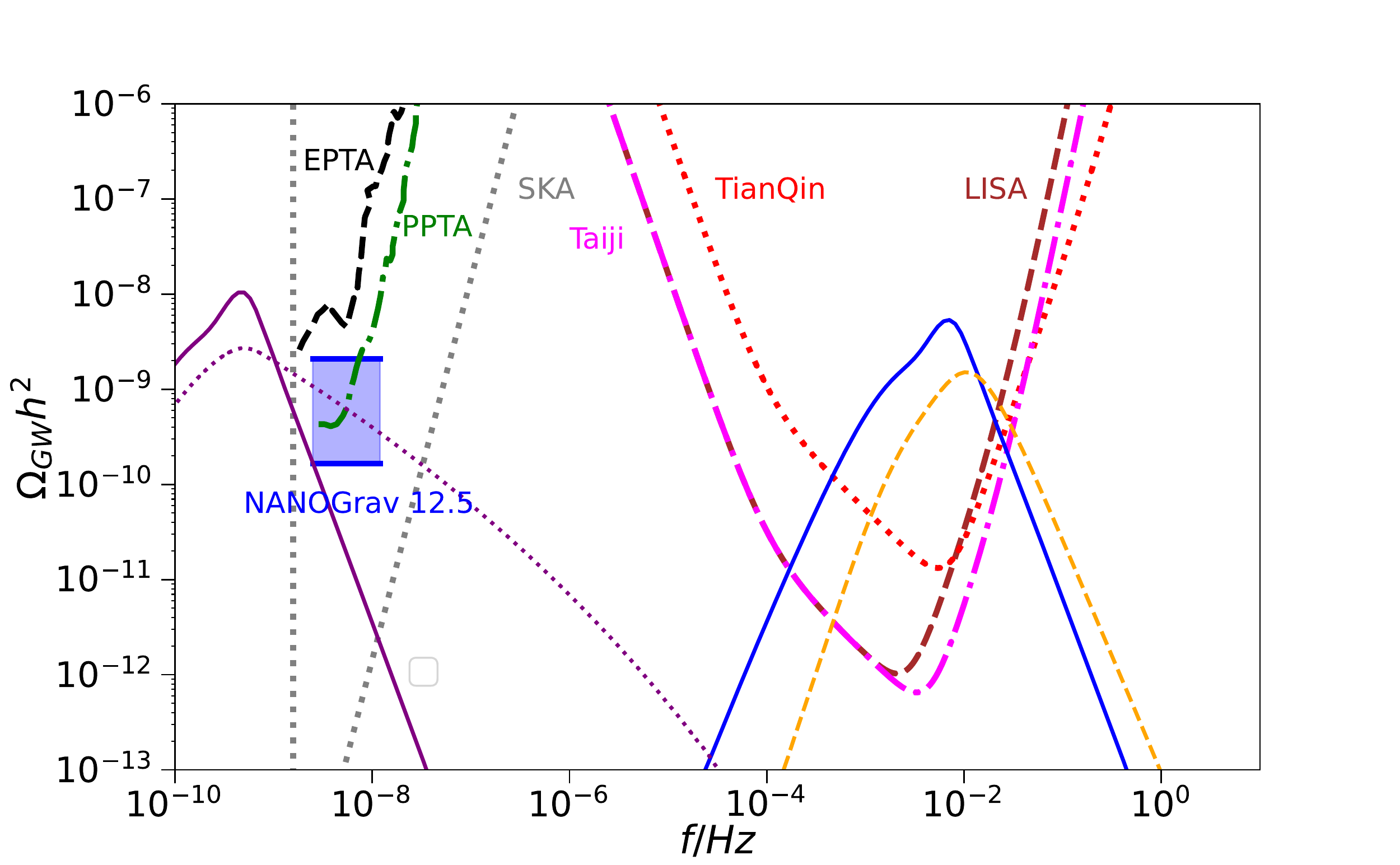}
\caption{The SIGWs from
Model H1 (the purple line), Model H2 (the blue line), WH (the purple  dotted line) and Q (the orange dashed line). 
The black dashed curve shows the EPTA limit \cite{Ferdman:2010xq,Hobbs:2009yy,McLaughlin:2013ira,Hobbs:2013aka} ,
the grey dotted curve denotes the SKA limit \cite{Moore:2014lga},
the brown dashed curve shows the LISA limit \cite{Audley:2017drz},
the red dot-dashed curve denotes the TianQin limit \cite{Luo:2015ght} and the dotted magenta curve denotes the TaiJi limit \cite{Hu:2017mde}}
\label{Fig:sigw}
\end{figure}

Substituting the obtained power spectrum from Higgs inflation in Sec.\ref{sec:2a} into Eqs.\eqref{fpbheq1} and \eqref{gwres1}, we get the PBH abundances{\footnote{Here we use the Gaussian formulate Eq.\eqref{betaG} for the fraction energy density of  PBHs at the formation.
Next section the non-Gaussianity effect will be taken into account.}}
as shown in Table.\ref{tab:2} and Fig.\ref{Fig:pbh} and the current energy densities of SIGWs as shown in Fig.\ref{Fig:sigw}.
 Model H1  produces PBHs with mass
 $M \simeq 30 M_{\odot}$ and the abundance $Y^{\text{peak}} \simeq 5.3\times10^{-4}$, which may explain the BH event GW150914 observed by LIGO \cite{Abbott:2016blz}. 
 The accompanying SIGWs have the peak frequency $f\sim 4.8\times10^{-10}\rm{Hz}$ and could be tested by SKA. 
 Although Model WH can not produce significant PBHs using the Gaussian formulate Eq.\eqref{betaG} ,  
 the energy density of SIGWs lies within the $2\sigma$ region of the NANOGrav signal \cite{DeLuca:2020agl,Inomata:2020xad,Vaskonen:2020lbd,Kohri:2020qqd,Domenech:2020ers,Vagnozzi:2020gtf,Kawasaki:2021ycf}.
Thus NANOGrav signal may originate from the Higgs field. 
Models H2  produces PBHs with mass $M \simeq\mathcal{O}( 10^{-13}) M_{\odot}$. 
In these mass ranges, PBHs can constitute almost all DM.
The accompanying SIGWs has the millihertz frequency, which can be tested by future space-based detectors like LISA, TaiJi, and TianQin.

\begin{table*}[htp]
  \centering
  \begin{tabular}{cccccccc}
  \hline
  Model &$k_{\text{peak}}/\text{Mpc}^{-1}$&$\mathcal{P}_{\zeta(\text{peak})}$&$M_{\text{PBH}}^{\text{peak}}/M_{\odot}$&$Y_{\text{PBH}(\text{peak})}^G$&$f_c$/Hz\\
  \hline
  H1 &$2.84\times10^5$&0.036&29&$5.3\times10^{-4}$&$4.8\times10^{-10}$\\
 H2  &$4\times10^{12}$&0.0258&$1.48\times10^{-13}$&0.88&$6.8\times10^{-3}$\\
 WH &$2.99\times10^{5}$&0.016&&&$5\times10^{-10}$\\
  Q &$6.53\times10^{12}$&0.013&&&$1.1\times10^{-2}$\\
  \hline
  \end{tabular}
 \caption{The results for PBH abundance and critical frequency of SIGWs.}
\label{tab:2}
\end{table*}

\section{Primordial non-Gaussianity}\label{sec:ng}
Due to the violation of the slow-roll condition, the non-Gaussianity may be large. 
Thus it is necessary to investigate the impact of non-Gaussianity on PBH abundance and SIGWs. In this section, we first compute the primordial non-Gaussianity from our model and then we discuss its effects on PBH abundance and the energy density of SIGWs.

The non-Gaussianity parameter $f_\text{NL}$ is \cite{Byrnes:2010ft}
\begin{equation}\label{Fnl}
f_{\text{NL}}(k_1,k_2,k_3)=\frac{5}{6}\frac{B_{\zeta}(k_1,k_2,k_3)}{P_{\zeta}(k_1)
P_{\zeta}(k_2)+P_{\zeta}(k_2)P_{\zeta}(k_3)+P_{\zeta}(k_3)P_{\zeta}(k_1)},
\end{equation}
where $P_{\zeta}(k)=2\pi^2\mathcal{P}_{\zeta}(k)/k^3$ and the bispectrum is defined as
\begin{equation}
\left\langle\hat{\zeta}_{\bm{k}_{1}}\hat{\zeta}_{\bm{k}_{2}}\hat{\zeta}_{\bm{k}_{3}}\right\rangle=(2 \pi)^{3} \delta^{3}\left(\bm{k}_{1}+\bm{k}_{2}+\bm{k}_{3}\right) B_{\zeta}\left(k_{1}, k_{2}, k_{3}\right).
\end{equation}
The expression for bispectrum is presented in Appendix.\ref{app.A}.

\begin{figure}[htp]
\centering
\includegraphics[width=0.7\textwidth]{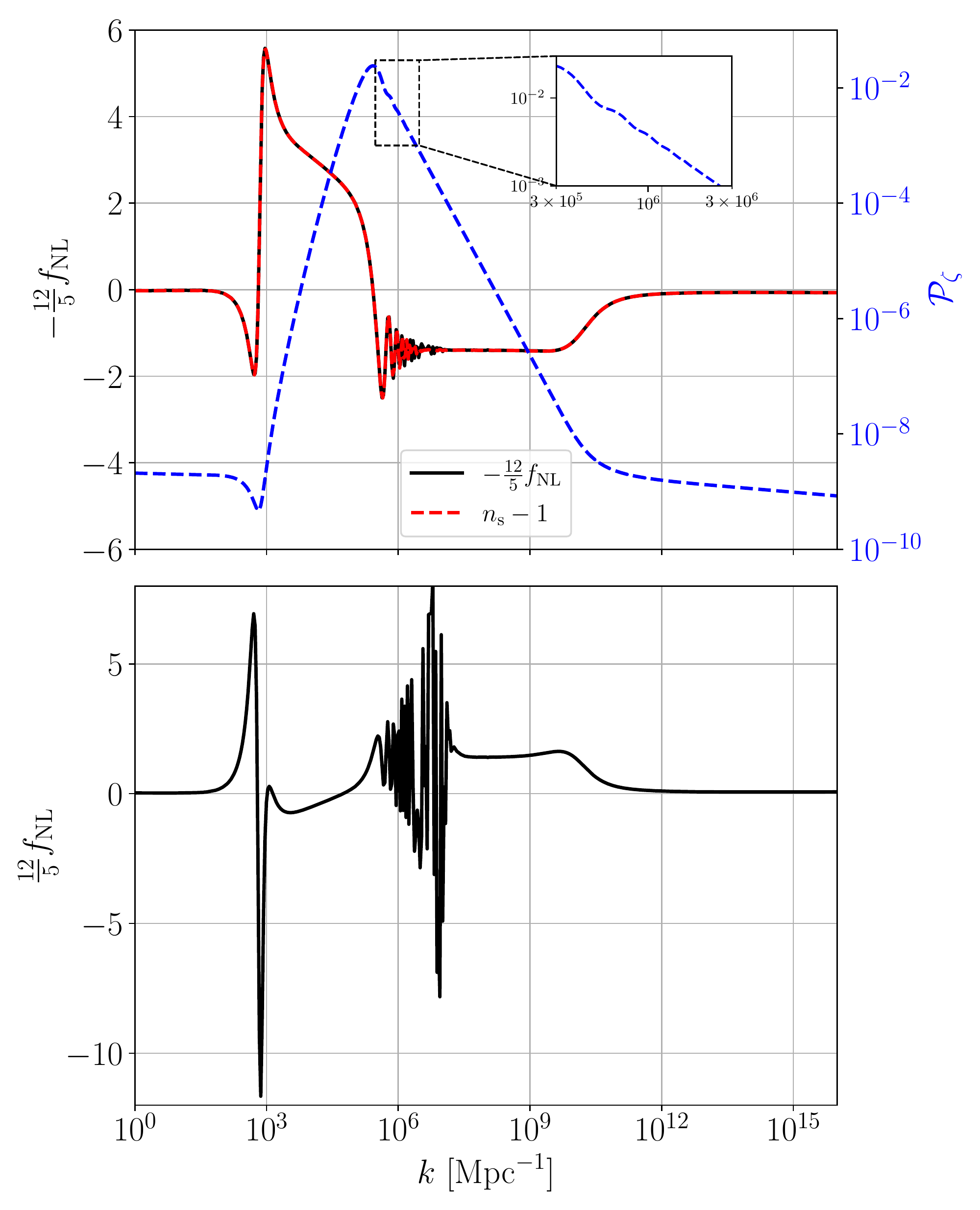}
\caption{The power spectrum of primordial curvature perturbation and the non-Gaussianity parameter $f_{\mathrm{NL}}$ for the model H1. 
We show the power spectrum with the blue dashed line, $-12f_{\mathrm{NL}}/5$ in the squeezed limit with the solid black line for the modes $k_1=k_2=10^6 k_3=k$ and the scalar spectral tilt $n_\mathrm{s}-1$ with the red dashed line in the upper panel.
The insets show the oscillations in $\mathcal{P}_\zeta$. The lower panel shows $12f_{\mathrm{NL}}/5$ in the equilateral limit for the modes $k_1=k_2=k_3=k$.}
\label{Fig:NH1}
\end{figure}

\begin{figure}[htp]
\centering
\includegraphics[width=0.7\textwidth]{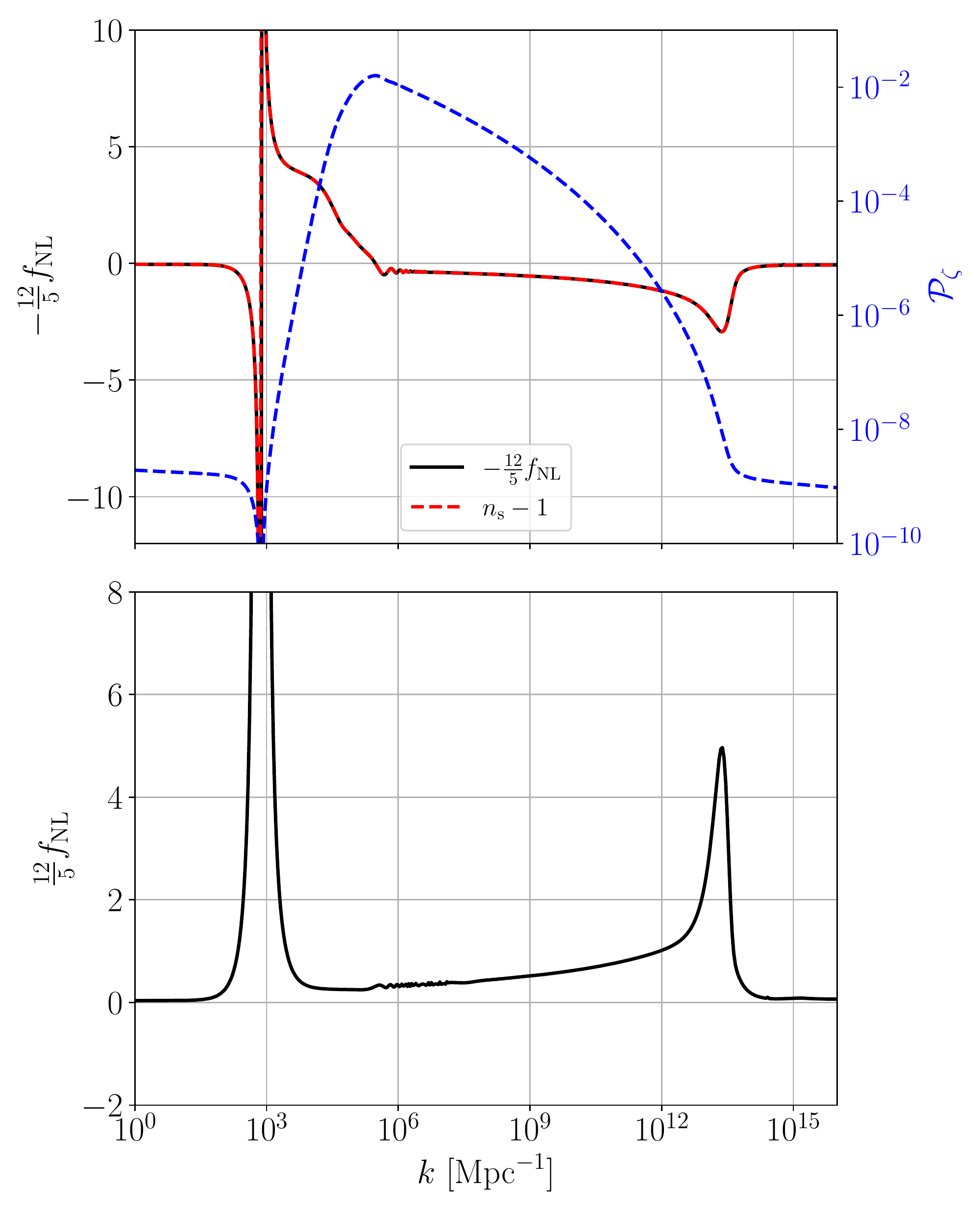}
\caption{The power spectrum of primordial curvature perturbation and the non-Gaussianity parameter $f_{\mathrm{NL}}$ for the model WH. 
We show $-12f_{\mathrm{NL}}/5$ in the squzeed limit with the solid black line for the modes $k_1=k_2=10^6 k_3=k$ and $12f_{\mathrm{NL}}/5$ in the equilateral limit for the modes $k_1=k_2=k_3=k$ in the upper and lower panels, respectively. We also show the power spectrum with the blue dashed line and the scalar spectral tilt $n_\mathrm{s}-1$ with the red dashed line in the upper panel.}
\label{Fig:NWH}
\end{figure}

With the parameter sets in table \ref{tab:1}, we numerically compute the non-Gaussianity parameter $f_{\text{NL}}$ and the results are shown in Figs. \ref{Fig:NH1} and \ref{Fig:NWH}. 
For single-field inflation, there is a consistency relation \cite{Maldacena:2002vr,Creminelli:2004yq} that relates the bispectrum and power spectrum,
\begin{equation}\label{consistency}
    \lim_{k_3\rightarrow 0} f_{\mathrm{NL}}(k_1,k_2,k_3)=\frac{5}{12}(1-n_{\mathrm s}),\quad \text{for}~ k_1=k_2,
\end{equation}
which can be used to test our numerical computation of non-Gaussianity. From Figs. \ref{Fig:NH1} and \ref{Fig:NWH}, we can see that the spectrum index $n_s-1$ matches with $-12f_{\text{NL}}/5$ in squeezed limit.
Note that from the expressions for bispectrum eqs. \eqref{etap}\eqref{eta}, the large non-Gaussianity parameter $f_{\text{NL}}$ may originate from the dramatic change in the velocity and acceleration of the inflaton.  
Note that the second slow-roll parameter $\epsilon_2$ for H1 ($q=1$, sharp power spectrum)  changes more dramatically than WH ($q=6/5$, broad power spectrum) around the peak scale.
Thus the non-Gaussianity parameter of WH around the peak region always stays smaller than H1, as shown in Table.\ref{tab:3}.

\subsection{Non-Gaussian effect on PBH abundance}
Now let us consider the non-Gaussian effect on PBH abundance.
Taking the non-Gaussianity correction into account, the fraction energy density of PBH at the formation becomes  \cite{Franciolini:2018vbk,Kehagias:2019eil,Atal:2018neu,Riccardi:2021rlf}
\begin{equation}
 \beta=\text{e}^{\Delta_3}\beta^{G}.
\end{equation}
The mass of PBHs we consider is almost monochromatic, thus the third cumulant $\Delta_3$ can be approximately expressed as \cite{Zhang:2021vak} 
\footnote{The sign of $\Delta_3$ differs from that in \cite{Zhang:2021vak}, which is incorrect. See the erratum of \cite{Zhang:2021vak}. }
\begin{equation}
\begin{split}
\Delta_3\approx  23\frac{\delta^3_c}{\mathcal{P}_{\zeta}(k_{\text{peak}})}f_{\mathrm{NL}}(k_{\text{peak}},k_{\text{peak}},k_{\text{peak}}).
\end{split}
\end{equation}
In Table.\ref{tab:3}, we show the results for the non-Gaussianity parameter $f_\text{NL}$ and the third cumulant $\Delta_3$.
The non-Gaussianity parameter $f_\text{NL}$ is of order $\mathcal{O}(10^{-1})$. 
In fact, during inflation with the sharper peak, the velocity of the inflaton  changes more dramatically such that the non-Gaussianity effect is more significant than inflation with the broad peak. 
The non-Gaussianity correction has a significant enhancement 
on PBH abundance, which means the formation of PBHs is easier with the consideration of the non-Gaussianity. 
The PBH abundance is underestimated with Gaussian statistics. 
Note that in Ref.\cite{Zhang:2020uek}, we have used the approximation formula for non-Gaussian PBH abundance proposed in  Ref. \cite{Saito:2008em} and concluded that for K/G inflation, the non-Gaussian effect on PBH abundance can be neglected due to $J_{\text{peak}}\sim \mathcal{O}(10^{-3})\ll1$.
However, this analysis neglected the factor before $J_{\text{peak}}$, i.e. $\left(\hat{\zeta}^5-8 \hat{\zeta}^3+9 \hat{\zeta}\right) J_{\text {peak }}$, where $\hat{\zeta} \equiv \zeta / \sqrt{\mathcal{P}_\zeta\left(k_{\text {peak }}\right)}$ and $J \equiv  \left\langle\zeta_{R_M}^3\right\rangle_c/ \sigma_{R_M}^3/6$.
Taking the critical value $\zeta_{c}\sim\mathcal{O}(1)$ \cite{Musco:2012au,Harada:2013epa,Young:2014ana,Motohashi:2017kbs,Franciolini:2018vbk} and thus $\hat{\zeta}_{c}\sim\mathcal{O}(10)$ , then we can find
$\left(\hat{\zeta}_c^5-8 \hat{\zeta}_c^3+9 \hat{\zeta}_c\right) J_{\text {peak}}\sim\mathcal{O}(10^2)$, which is large enough so that the non-Gaussian effect on PBH cannot be neglected.
On the other hand, note that the formula for non-Gaussian PBH abundance in Ref.\cite{Saito:2008em}  is an approximation result requiring the term $(\text{prefactor} \times J) \ll 1$, which is not satisfied in K/G inflation. 
Thus in this paper and Ref.\cite{Zhang:2021vak}, we adopt the exact formula for non-Gaussian PBH abundance proposed in Ref.\cite{Franciolini:2018vbk}.

\begin{table*}[htp]
  \centering
  \begin{tabular}{ccc}
  \hline
  Model & $f_\text{NL}(k_{\text{peak}},k_{\text{peak}},k_{\text{peak}})$ & $\Delta_3$\\
  \hline
  H1 &0.62&25\\
 H2  &0.53&30\\
WH & 0.13 &12\\
  \hline
  \end{tabular}
    \caption{The results for the non-Gaussianity parameter $f_\text{NL}$ and the third cumulant $\Delta_3$.}
\label{tab:3}
\end{table*}
\subsection{Non-Gaussian effect on SIGW}
To investigate the non-Gaussian effect on SIGWs, we first consider the power spectrum with the non-Gaussian correction.
The comoving curvature perturvation with the nonlinear corrections can be expressed as \cite{Verde:1999ij,Komatsu:2001rj}
\begin{equation}\label{local}
\zeta(\bm{x})=\zeta^G(\bm{x})+\frac{3}{5}f_\text{NL}(\zeta^G(\bm{x})^2-\langle \zeta^G(\bm{x})^2 \rangle),
\end{equation}
where $\zeta^G$ is the Gaussian part of the curvature perturbation.
Thus the power spectrum is 
\begin{equation}
\mathcal{P}_{\zeta}(k)=\mathcal{P}^G_{\zeta}(k)+\mathcal{P}^{NG}_{\zeta}(k),
\end{equation}
with the non-Gaussian correction of the power spectrum
\begin{equation}
\mathcal{P}^{NG}_{\zeta}(k)
=\left(\frac{3}{5}\right)^2\frac{k^3}{2\pi}f^2_{\mathrm{NL}}
\int d^3\bm{p}\frac{\mathcal{P}^G_{\zeta}(p)}{p^3}
\frac{\mathcal{P}^G_{\zeta}(|\bm{k}-\bm{p}|)}{|\bm{k}-\bm{p}|^3}.
\end{equation}    
For our model with $f_\text{NL}\sim\mathcal{O}(10^{-1})$ at peak scales, we have $\mathcal{P}^{NG}_{\zeta}(k)\ll\mathcal{P}^{G}_{\zeta}(k)$ and thus the non-Gaussian effect can be neglected when calculating SIGWs.

\section{Conclusions}\label{sec:conclu}
K/G inflation with a noncanonical kinetic term $(1+G(\phi))X$ can produce enhanced curvature perturbations at small scales if the coupling function $G(\phi)$ has a peak. 
However, due to the dramatic decrease in $\dot{\phi}$, 
the peak function $G(\phi)$ will contribute up to $\sim 20$ $e$-folds and the usual slow-roll inflation epoch endures around $30\text{-}40$ $e$-folds.
For power-law potential $V=\lambda\phi^p$, 
this indicates $p$ should be bounded as $p\lesssim 1$.
In particular, this mechanism does not work for Higgs potential $V=\lambda\phi^4$.
To resolve potential-restriction problem and produce PBHs and SIGWs in Higgs inflation, 
we introduce K/G inflation with nonminimal coupling and show that the curvature perturbation at small scales can be enhanced by the Higgs field while satisfying the constraints from CMB observations.
To be specific, in the Einstein frame, the conformal factor flattens the Higgs potential such that  the $e$-folds during slow-roll inflation is within $40$ and the tensor-to-scalar ratio is reduced. 

We then study the non-canonical kinetic coupling function in detail. 
On the one hand, phenomenologically, we find that the non-canonical kinetic coupling function not only drives the dramatic decreases in $\dot{\phi}$ and thus the enhancement of the curvature perturbation but also helps the exit of inflation. 
On the other hand, we study the observational constraints from LIGO merger rates and the White Dwarf explosion on the parameter space of the non-canonical kinetic coupling function. 
Our numerical results show that for a given width parameter $c$, there is an upper bound on the amplitude parameter $h$ and as $c$ gets larger, the upper bound on $h$ decreases.
The reason is that the larger $c$ indicates a wider peak function and thus the velocity of the inflaton decreases more dramatically and a smaller amplitude parameter $h$ is required to realize the same enhancement on curvature perturbations.

By varying the peak position, the curvature perturbation can be enhanced at different scales and thus different mass ranges of PBHs and frequencies of SIGWs can be produced. 
PBHs with mass $\mathcal{O}(10)M_{\odot},\ \mathcal{O}(10^{-12})M_{\odot}$ from models H1 and H2 may explain BHs in LIGO-Virgo events and almost all the DM, respectively. 
For SIGWs, we give the gauge invariant expression for the integral kernel of genuine SIGWs, which is related to terms propagating with the speed of light.  
The energy density of SIGWs from the model WH lies within the $2\sigma$ region of the NANOGrav signal.
Thus NANOGrav signal may originate from the Higgs field. 
SIGWs from the model H2 have the millihertz frequency, 
which can be tested by future space-based detectors like LISA, TaiJi, and TianQin.

Due to the violation of the slow-roll condition, the non-Gaussianity may have a significant effect on PBH abundance and SIGWs. 
Around the peak scale, we find that the non-Gaussianity parameter $f_\text{NL}$ of sharp power spectrum is larger than that of broad power spectrum  due to more dramatic change in the velocity of the inflaton. 
For our models, the non-Gaussianity parameter $f_\text{NL}$ in the equilateral limit is of order $\mathcal{O}(10^{-1})$ at peak scales and the non-Gaussianity correction has a significant enhancement on PBH abundance.  
Notwithstanding, the energy density of SIGWs remains invariant even if we take the non-Gaussianity into account, as the power spectrum receives very tiny corrections.

\begin{acknowledgments}
This research is supported in part by the National Natural Science Foundation of China under Grant No. 11875136,
and the Major Program of the National Natural Science Foundation of China under Grant No. 11690021.
J.L is also supported by the National Natural Science Foundation of China under Grant No.12247103, No.12047502 and No.12247117.
Y.L is also supported by the China Postdoctoral Science Foundation under Grant No. 2022TQ0140.

\end{acknowledgments}

\appendix
\section{The expression for the bispectrum}
\label{app.A}

The bispectrum $B_\zeta(k_1,k_2,k_3)$ is \cite{Hazra:2012yn,Ragavendra:2020old,Arroja:2011yj}
\begin{equation}
\label{bkeq1}
B_{\zeta}(k_1,k_2,k_3)=\Sigma^{10}_{i=1}B^{i}_{\zeta}(k_1,k_2,k_3),
\end{equation}
\begin{equation}\label{B1}
\begin{split}
B^{1}_{\zeta}(k_1,k_2,k_3)=&
-4\operatorname{Im}\left[\zeta_{{k}_{1}}(\tau_*)\zeta_{{k}_{2}}(\tau_*) \zeta_{{k}_{3}}(\tau_*) \int_{\tau_i}^{\tau_{*}}d\tau a^2\epsilon_1^{2}\left\{\zeta^{*}_{{k}_{1}}(\tau)\zeta^{'*}_{{k}_{2}}(\tau)\zeta^{'*}_{{k}_{3}}(\tau)
+\mathrm{perm}\right\}\right],
\end{split}
\end{equation}

\begin{equation}
\begin{split}
B^{2}_{\zeta}(k_1,k_2,k_3)=&
2\operatorname{Im}\left[
\vphantom{\left(\frac{\bm{k}_{2} \cdot \bm{k}_{3}}{k_{3}}\right)^{2}}
\zeta_{{k}_{1}}(\tau_*)\zeta_{{k}_{2}}(\tau_*) \zeta_{{k}_{3}}(\tau_*) \right.\\& \left. \int_{\tau_i}^{\tau_{*}}d\tau
a^2\epsilon_1^{2}\left\{(k^2_1-k^2_2-k^2_3)\zeta^{*}_{{k}_{1}}(\tau)\zeta^{*}_{{k}_{2}}(\tau)\zeta^{*}_{{k}_{3}}(\tau)
+\mathrm{perm}\right\}\right],
\end{split}
\end{equation}

\begin{equation}
\begin{split}
B^{3}_{\zeta}(k_1,k_2,k_3)=&
2\operatorname{Im}\left[
\vphantom{\left(\frac{\bm{k}_{2} \cdot \bm{k}_{3}}{k_{3}}\right)^{2}}
\zeta_{{k}_{1}}(\tau_*)\zeta_{{k}_{2}}(\tau_*) \zeta_{{k}_{3}}(\tau_*) \right. \\& \left. \int_{\tau_i}^{\tau_{*}}d\tau
a^2\epsilon_1^{2}\left\{\left(\frac{k^2_2-k^2_1-k^2_3}{k^2_1}+\frac{k^2_1-k^2_2-k^2_3}{k^2_2}\right)
\zeta^{'*}_{{k}_{1}}(\tau)\zeta^{'*}_{{k}_{2}}(\tau)\zeta^{*}_{{k}_{3}}(\tau)+\mathrm{perm}
\right\}\right],
\end{split}
\end{equation}

\begin{equation}\label{etap}
B^{4}_{\zeta}(k_1,k_2,k_3)=
-2\operatorname{Im}\left[\zeta_{{k}_{1}}(\tau_*)\zeta_{{k}_{2}}(\tau_*) \zeta_{{k}_{3}}(\tau_*) \int_{\tau_i}^{\tau_{*}}d\tau a^2\epsilon_1 \eta'\left\{\zeta^{*}_{{k}_{1}}(\tau)\zeta^{*}_{{k}_{2}}(\tau)\zeta^{'*}_{{k}_{3}}(\tau)+\mathrm{perm}\right\}\right],
\end{equation}

\begin{equation}
\begin{split}
B^{5}_{\zeta}(k_1,k_2,k_3)=&
-\frac{1}{2}\operatorname{Im}\left[
\vphantom{\left(\frac{\bm{k}_{2} \cdot \bm{k}_{3}}{k_{3}}\right)^{2}}
\zeta_{{k}_{1}}(\tau_*)\zeta_{{k}_{2}}(\tau_*)\zeta_{{k}_{3}}(\tau_*) \right.\\& \left.
\int_{\tau_i}^{\tau_{*}}d\tau a^2\epsilon_1^{3}\left\{\left(\frac{k^2_2-k^2_1-k^2_3}{k^2_1}+\frac{k^2_1-k^2_2-k^2_3}{k^2_2}\right)
\zeta^{'*}_{{k}_{1}}(\tau)\zeta^{'*}_{{k}_{2}}(\tau)\zeta^{*}_{{k}_{3}}(\tau)
+\mathrm{perm}\right\}\right],
\end{split}
\end{equation}

\begin{equation}
\begin{split}
B^{6}_{\zeta}(k_1,k_2,k_3)=&
-\frac{1}{2}\operatorname{Im}\left[
\vphantom{\left(\frac{\bm{k}_{2} \cdot \bm{k}_{3}}{k_{3}}\right)^{2}}
\zeta_{{k}_{1}}(\tau_*)\zeta_{{k}_{2}}(\tau_*) \zeta_{{k}_{3}}(\tau_*)\right. \\& \left.
\int_{\tau_i}^{\tau_{*}}d\tau a^2\epsilon_1^{3}\left\{\frac{k^2_3 \left(k^2_3-k^2_1-k^2_2\right)}{k^2_1k^2_2}
\zeta^{'*}_{{k}_{1}}(\tau)\zeta^{'*}_{{k}_{2}}(\tau)\zeta^{*}_{{k}_{3}}(\tau)
+\mathrm{perm}\right\}\right],
\end{split}
\end{equation}

\begin{equation}\label{eta}
B^{7}_{\zeta}(k_1,k_2,k_3)=2\operatorname{Im}\left[\zeta_{k_{1}}(\tau_*) \zeta_{k_{2}}(\tau_*) \zeta_{k_{3}}(\tau_*)
\left(a^{2} \epsilon_1 \eta\zeta_{k_{1}}^{*}(\tau) \zeta_{k_{2}}^{*}(\tau) \zeta_{k_{3}}^{\prime *}(\tau)+\mathrm{perm}\right)\right]\Big|_{\tau_i}^{\tau_*},
\end{equation}

\begin{equation}
\begin{split}
B^{8}_\zeta(k_1,k_2,k_3)=&2\operatorname{Im}\left[\zeta_{k_1}(\tau_*) \zeta_{k_2}(\tau_*) \zeta_{k_3}(\tau_*)\times\left(\frac{a}{H} \zeta_{k_{1}}^{*}(\tau) \zeta_{k_2}^{*}(\tau) \zeta_{k_{3}}^{*}(\tau)\right) \right.\\& \left.
\times \left\{54(a H)^{2}+2(1-\epsilon_1)(\bm{k}_1 \cdot \bm{k}_2+\mathrm{perm}) 
+ \right. \right. \\& \left. \left.  \frac{1}{2(a H)^{2}}\left[\left(\bm{k}_1 \cdot \bm{k}_2\right) k_{3}^{2}+\mathrm{perm}\right]\right\}\right] \Big|_{\tau_i}^{\tau_*},
\end{split}
\end{equation}

\begin{equation}
\begin{split}
B^{9}_{\zeta}(k_1,k_2,k_3)=&-\operatorname{Im}\left[\vphantom{\left(\frac{\bm{k}_{2} \cdot \bm{k}_{3}}{k_{3}}\right)^{2}} \zeta_{k_{1}}(\tau_*) \zeta_{k_{2}}(\tau_*) \zeta_{k_{3}}(\tau_*) 
 \left\{\frac{\epsilon_1}{ H^{2}} \zeta_{k_{1}}^{*}(\tau) \zeta_{k_{2}}^{*}(\tau) \zeta_{k_{3}}^{\prime *}(\tau) \right. \right. \\& \left.\left.
\left[k_{1}^{2}+k_{2}^{2}
-\left(\frac{\bm{k}_{1} \cdot \bm{k}_{3}}{k_{3}}\right)^{2}
-\left(\frac{\bm{k}_{2} \cdot \bm{k}_{3}}{k_{3}}\right)^{2}\right] +  \mathrm{perm}\right\}\right]\Big|^{\tau_*}_{\tau_i},
\end{split}
\end{equation}

\begin{equation}\label{B10}
\begin{split}
B^{10}_{\zeta}(k_1,k_2,k_3)=&2\operatorname{Im}\left[\vphantom{\left(\frac{\bm{k}_{2} \cdot \bm{k}_{3}}{k_{3}}\right)^{2}} \zeta_{k_{1}}(\tau_*) \zeta_{k_{2}}(\tau_*) \zeta_{k_{3}}(\tau_*) \right.\\
&\left. \left.
\times \left\{\frac{a\epsilon_1}{H} \zeta_{k_{1}}^{*}(\tau) \zeta_{k_{2}}^{\prime *}(\tau) \zeta_{k_{3}}^{\prime *}(\tau)
\left[2-\epsilon_1+\epsilon_1\left(\frac{\bm{k}_2\cdot \bm{k}_3}{k_2 k_3}\right)^2\right] +\mathrm{perm}\right\}\right]\right|^{\tau_*}_{\tau_i},
\end{split}
\end{equation}
where
$$
\eta=\dot{\epsilon}_1/H\epsilon_1,
$$
$\tau_i$ is the early time when all relevant modes are well within the horizon and the plane-wave initial condition is imposed.
$\tau_*$ is the late time when all relevant modes have been frozen.

%

\end{document}